%% file: main_leaplab.tex
\documentclass{article}
\usepackage{leaplab}

\input{config}

\definecolor{paleviolet}{HTML}{E1EEFC}
\definecolor{lightgrey}{RGB}{247, 247, 247}
\definecolor{darkgrey}{rgb}{0.5, 0.5, 0.5}
\definecolor{darkgreen}{rgb}{0, 0.5, 0}
\newenvironment{leapabstract}{
  \begin{tcolorbox}[
    colback=lightgrey,
    colframe=takeawayFrame,
    boxrule=0.8pt,
    arc=0pt,
    left=16pt,
    right=16pt,
    top=12pt,
    bottom=12pt,
    width=\textwidth,
    enlarge left by=0mm,
    before skip=10pt,
    after skip=10pt
  ]
}{
  \end{tcolorbox}
}

\begin{document}

\makeatletter
\def\icmldate#1{\gdef\@icmldate{#1}}
\icmldate{\today}
\makeatother

\makeatletter
\fancypagestyle{fancytitlepage}{
  \fancyhead{}
  \lhead{\includegraphics[height=0.7cm]{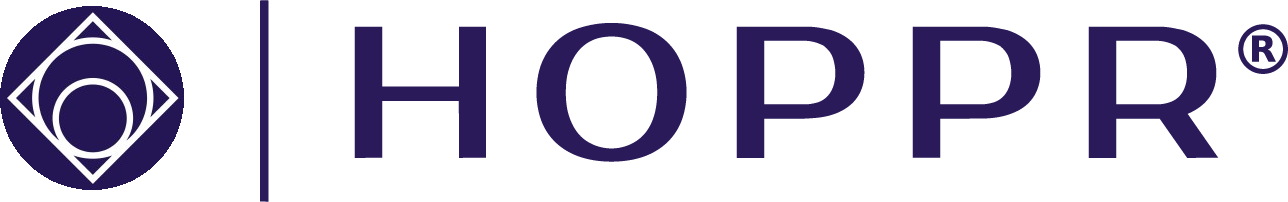}}
  \rhead{\it \@icmldate}
  \cfoot{}
}
\makeatother

\begin{center}
\icmltitle{Reconfigurable Radiology Labels Without Relabeling}
\icmltitlerunning{Reconfigurable Radiology Labels Without Relabeling}
\end{center}

\makebox[\textwidth][c]{%
\begin{tabular}{ccc}
\multicolumn{3}{c}{Jean-Benoit Delbrouck\;\; \qquad Dave Van Veen\;\;} \\[0.5em]
Akash Pattnaik\;\; &
Kalina Slavkova\;\; &
Javid Abderezaei\;\; \\[0.5em]
\multicolumn{3}{c}{Harris Bergman\;\; \qquad 
Khan Siddiqui\;\;}
\end{tabular}%
}


\begin{center}
HOPPR \\[0.25em]
\href{https://github.com/jbdel/radlabels}
{\raisebox{-0.15ex}{\small\faGithub}~\small github.com/jbdel/radlabels}
\end{center}

\icmlcorrespondingauthor{Jean-Benoit Delbrouck}{jeanbenoit.delbrouck@hoppr.ai}

{\let\thefootnote\relax\footnotetext{\hspace*{-\footnotesep}Correspondence to: Jean-Benoit Delbrouck \textless{}jeanbenoit.delbrouck@hoppr.ai\textgreater{}.}}

\vskip .3in

\begin{leapabstract}
    \input{tex/0-abs}
\end{leapabstract}

\input{tex/1-intro}
\input{tex/2-problem}

\input{tex/3-method}
\input{tex/4-experiments}
\input{tex/7-conclusion}

\newpage

\bibliography{main}
\bibliographystyle{icml2025}


\newpage
\appendix
\input{tex/5-appendix}


\end{document}

%% file: config.tex
\usepackage[table,x11names]{xcolor}

\usepackage[utf8]{inputenc} 
\usepackage[T1]{fontenc}    
\usepackage{url}            
\usepackage{booktabs}       
\usepackage{amsfonts}       
\usepackage{nicefrac}       
\usepackage{microtype}      
\usepackage{tabularx}   
\usepackage{enumitem}

\usepackage[utf8]{inputenc} 
\usepackage[T1]{fontenc}    
\usepackage{url}            
\usepackage{booktabs}       
\usepackage{amsfonts}       
\usepackage{nicefrac}       
\usepackage{microtype}      
\definecolor{citecolor}{HTML}{0071bc}
\definecolor{shadecolor}{HTML}{EFEFEF}
\definecolor{linkcolor}{HTML}{9A4D92}
\usepackage[colorlinks, linkcolor=linkcolor, colorlinks, anchorcolor=blue, citecolor=citecolor, pagebackref=True]{hyperref}

\usepackage{amsmath}
\usepackage{microtype}
\usepackage{graphicx}
\usepackage{booktabs} 
\usepackage{longtable}
\usepackage{caption}
\usepackage{wrapfig}
\usepackage{float}

\usepackage{titletoc} 

\usepackage{subcaption}
\usepackage[export]{adjustbox}
\usepackage{footnote}
\usepackage{tablefootnote}
\usepackage{threeparttable}
\usepackage[capitalize,noabbrev]{cleveref}

\usepackage{tabularray}
\UseTblrLibrary{booktabs}
\usepackage{multirow}
\usepackage{multicol}
\usepackage{xspace}
\usepackage{mathrsfs}
\usepackage{pifont}
\usepackage{makecell}
\usepackage{marvosym} 
\usepackage{fontawesome5} 
\usepackage[most]{tcolorbox}

\definecolor{takeawayBg}{HTML}{F7FAFF}    
\definecolor{takeawayFrame}{HTML}{BA8EFF} 
\newcommand{\takeawaybox}[2][Takeaway]{%
  \begin{tcolorbox}[
    colback=takeawayBg,
    colframe=takeawayFrame,
    boxrule=0.8pt,
    arc=0pt,
    left=12pt,
    right=12pt,
    top=8pt,
    bottom=8pt,
    width=\linewidth,
    before skip=8pt,
    after skip=8pt
  ]
  \textbf{#1.} #2
  \end{tcolorbox}
}

%% file: tex/0-abs.tex

Public chest-radiograph (CXR) datasets are typically released with small, fixed label schemas such as CheXpert-14. However, the underlying free-text reports describe far more findings---and which findings matter depends on the task, site, and reader. We release a pipeline that converts free-text reports into multi-label matrices with a single structured annotation pass, and then reconfigures the label schema through dictionary edits rather than new inference passes---i.e., without re-parsing or relabeling the corpus. After this one-time pass, reconfiguring MIMIC-CXR (223K reports) from cached annotations takes 196 seconds with no API cost, compared to \$6.6K for an equivalent relabeling pass with Claude Opus 4.7. Using a 58-label taxonomy, we show that 43\% of CXR studies contain at least one finding outside CheXpert-14. Image probes trained on these labels match CheXpert-14 probes on shared targets while also reaching 0.78 AUROC on expert-reviewed long-tail labels that CheXpert-14 cannot represent. These results suggest a different unit of work for radiology labeling: once reports are structured, the label schema becomes a configuration to edit, not a corpus to relabel.

%% file: tex/1-intro.tex
\section{Introduction}
\label{sec:intro}


Radiology image models are only as useful as the labels used to train and evaluate them. For many chest-radiograph (CXR) classifiers, those labels are a multi-label vector: one entry per finding, marked resent, absent, or uncertain. Over the last decade, these vectors have usually been derived from radiology reports by automatic labelers. On CXR, the standard tools are rule-based systems such as NegBio~\citep{peng2018negbio} and CheXpert-NLP~\citep{irvin2019chexpert}, followed by BERT-based classifiers such as CheXbert~\citep{smit2020chexbert}; all applied to the 14-finding CheXpert taxonomy. The same pattern appears beyond CXR: CT-RATE~\citep{hamamci2024ctrate} labels chest CT reports by annotating a small subset and distilling a RadBERT-style classifier to the remaining corpus. More recently, large language models (LLMs) have been used to extract broader label sets~\citep{dorfner2025llm}, while entity-relation extractors such as RadGraph~\citep{jain2021radgraph,delbrouck2024radgraph} produce structured report annotations that downstream tools can consume.
 

These developments have made large-scale radiology labeling practical, but most rely on fixed schemas. That is a lossy abstraction: reports describe devices, diseases, subtypes, uncertainty, anatomy, and site-specific phrasing that may not fit the schema. The right schema also depends on the task: one application may need pleural abnormality, another pleural effusion versus pleural thickening, and another laterality, severity, or temporal change. Beyond task dependencies, schema choices are also reader-dependent: radiologists may disagree on the presence or description of a finding~\citep{abujudeh2010abdominal,bruno2015understanding}.
These sources of variation make flexibility a useful property of a labeling system; however, fixed taxonomies constrain what models learn and what researchers can measure.

The practical barrier is schema iteration at corpus scale. Manual relabeling requires scarce radiologist time~\citep{mcdonald2015effects}. Distillation pipelines, such as the one used for CT-RATE, require a new annotation-and-training cycle when target labels change. LLMs can relabel reports more flexibly, but each schema change requires another inference pass over the corpus, introducing recurring cost, privacy constraints, and reproducibility concerns: one Claude Opus~4.7 pass over MIMIC-CXR costs approximately \$6.6K\footnote{At 2026-04 public pricing, assuming 162 input tokens/report (measured with tiktoken cl100k on a random sample of 1,000 reports), 800 tokens of prompt overhead, and 200 tokens of structured-JSON output.}, must send report text to an API unless run locally, and may change with the prompt or model. The bottleneck is therefore not producing labels once, but changing them safely, cheaply, and repeatedly.


\textbf{To address this, we propose a method for reconfiguring radiology label schemas without a repeated full-corpus inference pass: a single, one-time structured report annotation pass separates report understanding from downstream label definition.}
We use ``without relabeling'' throughout in this specific sense---the corpus is parsed once and never re-parsed---rather than to claim that no annotation ever occurs.
For each finding, such as pleural effusion, a radiologist defines a small set of phrases that express the finding in reports, such as ``pleural effusion'', ``hydrothorax'', or ``fluid in the pleural space''.
A Structured Report Annotator (SRA) runs once over the corpus and extracts, for each report, the findings described by the report author, each marked as present, uncertain, or absent.
Labels are then compiled by matching the radiologist-defined phrases against this cached structured output. The alias dictionary, which we call Radiological Aliases, maps each target label to the report phrases that express it.
Reconfiguring the schema---editing a label, adding a new one, or adapting labels to a new institution---requires only a dictionary edit and local recompile; the structured annotation pass is not re-run, so no additional full-corpus inference is needed.
Each label also stores the matched phrase and report tokens, so a reviewer can inspect the evidence that produced it.
Our code and the full alias dictionary are released at \url{https://github.com/jbdel/radlabels}.


The rest of the paper is organized as follows.
\Cref{sec:problem} defines the labeling problem and motivates the need for reconfigurable schemas.
\Cref{sec:method} describes the two components of our approach: the SRA (\Cref{sec:sra}) and Radiological Aliases (\Cref{sec:aliases}).
\Cref{sec:exp} describes the datasets, taxonomy, and evaluation setup (\Cref{sec:resources}), then evaluates report-label extraction (\Cref{sec:results}) and downstream image-model utility (\Cref{sec:training-results}).
\Cref{sec:limitations} discusses limitations, and \Cref{sec:conclusion} concludes.

%% file: tex/2-problem.tex
\section{Problem Statement}
\label{sec:problem}

Public radiological datasets ship with a small, fixed set of labels, usually CheXpert-14~\citep{irvin2019chexpert} on CXR or the 18-finding panel distilled from radiologist annotations on CT-RATE~\citep{hamamci2024ctrate} on chest CT.
Fixed sets are convenient for benchmarking, but they fit reports poorly.
The bottleneck is not only producing labels; it is defining, changing, and auditing them.

\takeawaybox{Radiology label schemas are clinical choices, not fixed objects; they change with disease, reader, site, and task.}

\paragraph{(i) Labels are open-ended.}
New findings, devices, and subtypes appear over time: LVADs, Impella pumps, and newer pacemaker models are not in any of the taxonomies above.
Reports are also long-tailed: common findings dominate, while many clinically important findings each appear in under 1\% of studies~\citep{holste2022long}.
Even on CXR alone, PadChest~\citep{bustos2020padchest} contains 174 distinct radiographic findings, which is more than ten times CheXpert-14.
Even for a fixed disease, the ``right'' label depends on downstream use: for one study one may want coarse \textsc{Pleural Abnormality}; for another, fine-grained \textsc{Pleural Thickening} vs \textsc{Pleural Effusion} vs \textsc{Pneumothorax}; for a third, a laterality- or severity-aware version.
This mismatch is not just aesthetic: hidden subgroups inside coarse labels cause clinically meaningful failures in downstream models~\citep{oakden2020hidden}.

\paragraph{(ii) Radiologists disagree with one another.}
Inter-observer variability on CXR and cross-sectional imaging is well documented: discrepancy rates between experienced readers run in the 2--20\% range depending on the task~\citep{abujudeh2010abdominal,bruno2015understanding}.
The same finding may also be phrased as ``effusion'', ``fluid in the pleural space'', or ``blunting of the costophrenic angle''.
A fixed schema hard-codes one group's choices and makes every downstream user inherit them.

\paragraph{(iii) Manual relabeling does not scale.}
Radiologists are an increasingly scarce resource: workload per radiologist has grown substantially with the volume of cross-sectional imaging~\citep{mcdonald2015effects}.
Relabeling a single large dataset like MIMIC-CXR (223k reports) with new labels would occupy a radiologist full-time for many months, and the process must be repeated each time the schema changes.
When a hand-annotated subset exists, the usual workaround is distillation: CT-RATE~\citep{hamamci2024ctrate}, for example, involved annotating 18 findings on a small subset and subsequently training a RadBERT-style classifier to label the rest.
Changing those 18 findings still requires another annotation-and-distillation cycle.
The exceptional hand-curated datasets we rely on for evaluation --CheXpert~5x200~\citep{irvin2019chexpert}, CXR-LT~2024 Task~2~\citep{holste2024cxrlt}, and PadChest-GR~\citep{castro2024padchestgr} -- exist precisely because they are small (hundreds to a few thousand studies) or because large consortia were organised around them.

\paragraph{(iv) LLM relabeling is fast but costly, privacy-constrained, and non-deterministic.}
LLMs can in principle label anything a radiologist can label, and recent work has shown that LLMs are competitive with expert labelers for CXR reports~\citep{dorfner2025llm}.
At scale, however, cost, privacy, speed, and reproducibility all become problems.
A MIMIC-scale corpus relabeled with a frontier model costs about \$6.6k and scales linearly with corpus size (\Cref{sec:cost}).
Many institutions cannot send report text to third-party APIs; batch jobs can take days; and prompt or model changes silently shift the labels.
Rule-based and BERT-based labelers avoid cost and privacy issues but hard-code a fixed taxonomy and are also known to introduce their own noise~\citep{olatunji2019caveats}.

\takeawaybox{The bottleneck is not producing labels once; it is changing them safely, cheaply, and repeatedly.}

\paragraph{Desired result.}
A useful pipeline should let users \emph{reconfigure} the label schema without relabeling the corpus: changes should be cheap to apply, local and deterministic, open-ended, and auditable down to the report tokens that produced each label.
Our method, described next, satisfies all four.

%% file: tex/3-method.tex
\section{Method}
\label{sec:method}

Our method has two components, mirroring the high-level picture from the introduction.
A \emph{Structured Report Annotator} (\Cref{sec:sra}) reads each report once and turns it into a structured list of findings with statuses.
\emph{Radiological Aliases} (\Cref{sec:aliases}) is a small, radiologist-edited dictionary that maps each label to a few phrases. Matching those phrases against the parsed output produces the final labels.
The annotator pass is the only expensive step and is cached so that changing the label set is just a dictionary edit.

\subsection{Structured Report Annotators (SRAs)}
\label{sec:sra}

A \textbf{Structured Report Annotator (SRA)} turns a radiology report into a \emph{semantic graph}: nodes are clinical concepts, edges are relationships, and each node carries a status (present, absent, or uncertain).

In this work, we use RadGraph-XL~\citep{delbrouck2024radgraph} as the SRA.
Its nodes are \texttt{Anatomy} (e.g.\ ``pleura'', ``aorta'') or \texttt{Observation} (e.g.\ ``effusion'', ``pacemaker''), each tagged as \texttt{definitely present}, \texttt{uncertain}, or \texttt{definitely absent}.
Its edges encode relations such as \texttt{located\_at} (\emph{effusion} $\xrightarrow{\texttt{located\_at}}$ \emph{pleural}), \texttt{modify} (\emph{small} $\xrightarrow{\texttt{modify}}$ \emph{effusion}), and \texttt{suggestive\_of}.
Each node also stores token positions so that every label can point back to report text.
Other SRAs fit the same mould\footnote{For example, Google's \emph{langextract}~\citep{goel2023langextract} can be prompted with a radiology schema to emit comparable annotations; such LLM-based pipelines become practical at full-corpus scale once distilled into a smaller open-source model, which is essentially how RadGraph itself was built.} and are drop-in replacements for RadGraph-XL in this proposed framework.

Running the SRA is the only computationally expensive step, and it is paid once per report.
Annotations are cached, and downstream label changes only necessitate recompiling aliases.
With data-parallel inference, our setup processes 100k reports in about 22 minutes on 3 A100s; \Cref{sec:cost} reports end-to-end throughput.

\takeawaybox{Run the Structured Report Annotator once; reuse its structured output for future label definitions.}

\subsection{Radiological Aliases}
\label{sec:aliases}

\textbf{Radiological Aliases} turns SRA annotations into a multi-label matrix.
The user edits a dictionary mapping each disease key to the following:
\begin{itemize}[itemsep=1pt, topsep=2pt, leftmargin=*]
\item a list of \textbf{alias phrases} (e.g.\ \texttt{pleural\_effusion}: ``pleural effusion'', ``pleural fluid'', ``hydrothorax'', \ldots);
\item an optional \texttt{parent} (e.g.\ \texttt{pleural\_abnormality}) derived from child labels and used in the downstream experiments (\Cref{sec:downstream});
\item optional \texttt{exclude} phrases for token-overlap collisions. For example, \texttt{pleural\_effusion} excludes ``pericardial effusion'' so an effusion around the heart is not labeled as pleural.
\end{itemize}
The full alias dictionary is released with the code; \Cref{sec:alias-appendix} and \Cref{tab:alias-summary} give a compact view of the label groups and representative aliases used in this paper.

\paragraph{Labeling a study.}
For each observation node, we collect the words attached to it (anatomy, modifiers, and connected observations) and check whether any disease alias appears inside that cluster.
If they do, the disease is recorded with the observation's status, unless one of its \texttt{exclude} phrases also appears in the cluster, in which case we ignore the hit.
A disease with several hits keeps the strongest status (\emph{present} $>$ \emph{uncertain} $>$ \emph{absent}); parents inherit from children.
In downstream experiments, only \emph{present} labels are used as positives. An absent child does not by itself prove that all siblings are absent.
Implementation details are in \Cref{sec:algo-appendix}.

\paragraph{Worked example.}
Take ``\emph{acute nondisplaced 7th rib fracture, suggestive of recent trauma}''.
RadGraph returns the following structured annotation for this sentence:
\[
\resizebox{\linewidth}{!}{$
\text{acute} \xrightarrow{\texttt{modify}} \text{fracture} \xrightarrow{\texttt{located\_at}} \text{rib}, \quad
\text{nondisplaced} \xrightarrow{\texttt{modify}} \text{fracture}, \quad
\text{7th} \xrightarrow{\texttt{modify}} \text{rib}, \quad
\text{fracture} \xrightarrow{\texttt{suggestive\_of}} \text{trauma}
$}
\]
Here, \emph{fracture} is the present observation and \emph{rib} is the anatomy.
The cluster is $\{\text{acute, nondisplaced, 7th, fracture, rib, trauma}\}$.
Aliases such as ``acute rib fracture'' and ``nondisplaced rib fracture'' match, so \texttt{acute\_rib\_fracture} is labeled \emph{present}; its parent, \texttt{fracture\_or\_trauma}, inherits the status.
Each hit stores the matched alias and token positions, so the label traces back to the report.

\takeawaybox{A label is not just a column value; it is a matched clinical phrase tied to report evidence.}

%% file: tex/4-experiments.tex
\section{Experiments}
\label{sec:exp}

We evaluate two questions: can we extract labels cheaply and accurately, and do those labels help image models?
\Cref{sec:resources} defines the datasets, taxonomy, and method names.
\Cref{sec:results} evaluates label extraction.
\Cref{sec:training-results} evaluates training with extracted labels.

\subsection{Resources: datasets and taxonomy}
\label{sec:resources}

\paragraph{Datasets.}
We use three report corpora.
\emph{MIMIC-CXR}~\citep{johnson2019mimic} contributes 222{,}694 labeled studies (211{,}699 with image features in the classification experiments).
\emph{CheXpert+}~\citep{chambon2024chexpertplus} contributes 187{,}330 labeled studies.
\emph{HOPPR CXR}, an internal de-identified industry corpus sourced from the HOPPR Platform, contributes 455{,}382 labeled studies for the label-space audit, including an 86{,}049-study representative image subset for classification.
These de-identified studies are aggregated from data providers across eight U.S. states; each study in the image subset corresponds to a distinct patient, so study-level splitting induces no same-patient leakage.
Together, these corpora test whether one alias dictionary can run across different institutions and case mixes.

We evaluate label quality on three gold sets.
\emph{CheXpert 5x200} is a 1{,}000-study expert-reviewed subset with 200 positives for each of five CheXpert labels.
\emph{CXR-LT 2024 Task 2}~\citep{holste2024cxrlt} contains 407 expert-reviewed MIMIC-CXR studies with 26 fine-grained labels, many outside CheXpert-14.
We evaluate 25 non-normal report labels and 21 image labels with support in the feature/gold-label intersection.
\emph{PadChest-GR}~\citep{castro2024padchestgr} contains 4{,}555 studies with radiologist-reviewed grounded findings, translated from Spanish into English finding sentences.

Each gold set stresses a different property:
\begin{enumerate}[itemsep=1pt, topsep=2pt, leftmargin=*]
\item CheXpert 5x200 tests compatibility with the standard CheXpert labeler setting.
\item CXR-LT stresses long-tail CXR findings that are mostly absent from CheXpert-14.
\item PadChest-GR stresses vocabulary transfer across institution, language, and label convention.
\end{enumerate}

\paragraph{Taxonomy.}
All main experiments use a frozen 58-label taxonomy: 48 fine-grained leaves and 10 grouping labels.
The leaves cover CheXpert-14, CXR-LT 2024, and additional device, pleural, cardiomediastinal, fracture, airway, and rare-air findings.
The 10 grouping labels are \texttt{air\_space\_opacity}, \texttt{airway\_abnormality}, \texttt{cardiomediastinal\_abnormality}, \texttt{fracture\_or\_trauma}, \texttt{hernia\_abnormality}, \texttt{hilar\_abnormality}, \texttt{mediastinal\_or\_abdominal\_air}, \texttt{pleural\_abnormality}, \texttt{pulmonary\_abnormality}, and \texttt{support\_devices}.
The hierarchy is not strictly two-level: a grouping label may itself roll up into a broader one (for example, \texttt{air\_space\_opacity} groups \texttt{consolidation} and \texttt{pneumonia} while itself rolling up under \texttt{pulmonary\_abnormality}). For the classification experiments we use a single parent level per leaf; the full parent-of-parent structure is shown in \Cref{tab:full-label-mapping}.
CheXpert-14 includes \emph{No Finding}, but classifier heads exclude it, so CheXpert-style image experiments use 13 heads.
Grouping labels are derived from their children and used in the parent-conditioned classification experiment.
For PadChest-GR only, we additionally report a small synonym update that closes obvious vocabulary gaps (for example, PadChest-GR's ``aortic elongation'' maps to our \texttt{tortuous\_aorta}).
For clarity, we keep this update separate from the frozen-taxonomy result.
A small new-label demo is reported in \Cref{sec:control}.

Table names are as follows.
\emph{Radiological Aliases} is the full pipeline: SRA graph plus aliases.
\emph{Radiological Aliases (frozen)} uses the pre-specified taxonomy.
\emph{Radiological Aliases (PadChest synonyms)} adds transparent PadChest vocabulary synonyms.
\emph{Raw-text aliases} uses the same aliases but skips the SRA graph.

\subsection{Label extraction results}
\label{sec:results}

\paragraph{Reconfiguring from cached annotations costs minutes, not dollars.}
\label{sec:cost}
The SRA pass is paid only once per report.
After that, changing labels consists of a local recompile instead of reprocessing the corpus with an LLM every time the label schema changes.

\begin{table}[h]
\centering
\small
\caption{Estimated LLM relabeling cost (USD, input+output) versus Radiological Aliases recompile time from cached annotations. Costs use public prices from 2026-04 and repeat each time the label schema changes.}
\label{tab:cost}
\begin{tabular}{lrrrrrrr}
\toprule
Scale & Opus & Sonnet & Haiku & GPT-4-T & GPT-4o & 4o-mini & \textbf{Ours} \\
\midrule
100k          & \$2{,}943   & \$589   & \$157   & \$1{,}562  & \$440   & \$26  & \textbf{88s} \\
223k (MIMIC)  & \$6{,}553  & \$1{,}311 & \$350 & \$3{,}478  & \$981 & \$59 & \textbf{196s} \\
1.4M          & \$41{,}199  & \$8{,}240 & \$2{,}197 & \$21{,}866 & \$6{,}167 & \$370 & \textbf{1{,}229s} \\
5M            & \$147{,}140 & \$29{,}428 & \$7{,}847 & \$78{,}093 & \$22{,}023 & \$1{,}321 & \textbf{4{,}390s} \\
\bottomrule
\end{tabular}
\end{table}

On MIMIC-CXR scale, an Opus relabeling run is about \$6.6k; our recompile is about three minutes.
At 5M reports, Opus rises to roughly \$147k, while the local recompile remains under 75 minutes.
The LLM cost repeats whenever the label definition changes; our method edits aliases and recompiles.
Two costs are deliberately excluded from the ``Ours'' column and are not zero: the one-time SRA pass (compute reported in \Cref{sec:sra}, paid once per corpus) and the human labor of curating the alias dictionary. The recompile column measures the \emph{marginal} cost of a schema change once structured annotations exist, which is where our method is effectively free; it is not a claim that the end-to-end pipeline is free.

\takeawaybox{The expensive step is paid once; later schema changes are local recompiles, not new LLM labeling jobs.}

\paragraph{Closed-set labels miss a large fraction of what reports say.}
\label{sec:unlimited}
On 455{,}382 studies from HOPPR CXR, 43.0\% have at least one positive fine-grained label outside CheXpert-14.
Among study pairs with identical CheXpert-14 vectors, 52.3\% differ by at least one fine label, and 5.8\% differ by three or more.
The mean study has 0.99 fine-label positives, compared with 0.43 CheXpert-14 positives.

\takeawaybox{CheXpert-14 is useful, but it is a lossy compression of the report. Two studies can share the same coarse label vector while describing different clinical states.}

\paragraph{Positives are precise where vocabularies align.}
\label{sec:accuracy}
Radiological Aliases is precision-first: a positive fires only when a radiologist-defined phrase matches a structured annotation.
Because the method is closed-vocabulary, recall is bounded by alias coverage; we therefore report precision first and F1 for context, and we scope precision claims to the setting rather than claiming uniformly ``high precision.''
One caveat when reading the numbers below: our positives are scored against each gold set's own label \emph{definitions}. When two schemas draw category boundaries differently, a clinically reasonable positive can be counted as a false positive purely because of an ontology mismatch (for example, a finding we call \texttt{tortuous\_aorta} may be recorded under a different grounded label). Low precision against an external ontology is thus not always evidence of an incorrect label; it can reflect a definition gap that a dictionary edit can close (\Cref{tab:padchest}).
We compare against CheXbert~\citep{smit2020chexbert}, a strong fixed-vocabulary labeler, and against Raw-text aliases, which applies the same aliases directly to report text without the SRA graph.

On CheXpert 5x200---the standard CheXpert labeler setting---our precision matches CheXbert (0.97 vs 0.97).
Raw-text aliases are much less precise (0.74), showing why the SRA graph is useful even for familiar CheXpert labels.
On the broader long-tail panels, precision is lower but still substantial (macro-P 0.72 on CXR-LT, 0.66 on frozen PadChest-GR); we call this competitive rather than high precision.
The recall gap relative to CheXbert is expected: CheXbert is trained for its labels, while Radiological Aliases only fires when the dictionary covers the phrasing.

\begin{table}[h]
\centering
\small
\caption{CXR-LT 2024 Task 2. This is primarily a \emph{coverage} comparison, not a like-for-like accuracy comparison: CheXbert is a 14-output classifier and is structurally unable to emit 12 of the 25 evaluated labels. We therefore split results into the 13 CheXbert-representable labels and the 12 it cannot express.}
\label{tab:cxrlt}
\begin{tabular}{lccc}
\toprule
Method & macro-P & macro-R & macro-F1 \\
\midrule
\multicolumn{4}{l}{\emph{All 25 labels}} \\
\textbf{Ours} & \textbf{0.72 [0.61, 0.82]} & 0.60 [0.50, 0.68] & \textbf{0.61 [0.52, 0.69]} \\
Raw-text aliases & 0.60 [0.48, 0.72] & 0.55 [0.42, 0.65] & 0.50 [0.39, 0.59] \\
CheXbert & 0.33 [0.19, 0.48] & 0.40 [0.26, 0.55] & 0.34 [0.20, 0.49] \\
\midrule
\multicolumn{4}{l}{\emph{13 CheXbert-representable labels}} \\
\textbf{Ours} & \textbf{0.72 [0.55, 0.87]} & 0.53 [0.38, 0.66] & 0.54 [0.40, 0.66] \\
Raw-text aliases & 0.50 [0.32, 0.68] & 0.48 [0.32, 0.64] & 0.40 [0.27, 0.54] \\
CheXbert & 0.64 [0.48, 0.78] & 0.77 [0.70, 0.83] & 0.66 [0.53, 0.78] \\
\midrule
\multicolumn{4}{l}{\emph{12 non-CheXbert labels}} \\
\textbf{Ours} & \textbf{0.73 [0.59, 0.85]} & \textbf{0.68 [0.59, 0.76]} & \textbf{0.68 [0.58, 0.77]} \\
Raw-text aliases & 0.70 [0.56, 0.85] & 0.62 [0.44, 0.76] & 0.60 [0.45, 0.75] \\
CheXbert & 0.00 [0.00, 0.00] & 0.00 [0.00, 0.00] & 0.00 [0.00, 0.00] \\
\bottomrule
\end{tabular}
\end{table}

This comparison should be read as a coverage comparison, not a claim that our method is a more accurate labeler than CheXbert.
On the 13 labels CheXbert was built to output, \emph{CheXbert is more accurate than our method}: F1 0.66 vs 0.54, driven mainly by recall (0.77 vs 0.53) since our closed-vocabulary method only fires on covered phrasings, while our precision there is competitive (0.72 vs 0.64).
On the 12 labels CheXbert cannot represent, its F1 is 0.00 by construction; this is a statement about coverage, not accuracy, and we do not treat it as a head-to-head win.
The value of Radiological Aliases here is that a single dictionary spans the full 25-label panel at all, whereas a fixed 14-label head cannot; the honest summary is that our method trades some accuracy on CheXpert-representable labels for coverage of long-tail labels a fixed labeler cannot express.
The full per-label table is in \Cref{tab:cxrlt-full}.

\takeawaybox{A fixed labeler can be excellent inside its vocabulary and still blind outside it.}

\begin{table}[h]
\centering
\small
\caption{PadChest-GR accuracy. The same compiler runs on a different institution and a different label vocabulary. “Frozen” uses the original alias dictionary without PadChest-specific edits. “PadChest synonyms” adds dataset-specific synonym aliases to existing labels while keeping the same taxonomy, SRA output, and matching algorithm.}
\label{tab:padchest}
\begin{tabular}{lccc}
\toprule
Method & macro-P & macro-R & macro-F1 \\
\midrule
Radiological Aliases (frozen) & 0.66 [0.53, 0.78] & 0.45 [0.35, 0.56] & 0.50 [0.39, 0.61] \\
\textbf{Radiological Aliases (PadChest synonyms)} & \textbf{0.79 [0.67, 0.88]} & 0.57 [0.47, 0.67] & \textbf{0.62 [0.51, 0.71]} \\
Raw-text aliases (PadChest synonyms) & 0.73 [0.62, 0.84] & \textbf{0.63 [0.52, 0.73]} & \textbf{0.62 [0.51, 0.72]} \\
CheXbert & 0.00 [0.00, 0.00] & 0.00 [0.00, 0.00] & 0.00 [0.00, 0.00] \\
\bottomrule
\end{tabular}
\end{table}

PadChest-GR is a vocabulary stress test: many labels are outside CheXbert-14, and some use different names for findings we already represent.
The frozen row tests the original dictionary.
The PadChest-synonym row tests controllability: the added synonyms are taken from PadChest-GR's own controlled-vocabulary label \emph{definitions} (via a fixed vocabulary crosswalk, e.g.\ ``aortic elongation'' $\rightarrow$ \texttt{tortuous\_aorta}), not by inspecting gold test outcomes, so this is a vocabulary-alignment edit rather than test-set tuning.
Starting from the frozen dictionary, we add these synonyms to existing labels without changing the compiler or retraining, which raises macro-F1 from 0.50 to 0.62.
CheXbert remains at 0.00 because it cannot emit this vocabulary at all; this row should be read as ``not applicable outside CheXbert-14,'' not as a fair overlapping-vocabulary comparison.

\takeawaybox{Vocabulary mismatch is often a definition problem, not a model-retraining problem.}

\paragraph{The SRA graph matters.}
\label{sec:rq6}
This ablation asks whether the SRA graph is necessary.
Raw-text aliases use the same dictionary but search the report text directly.

\begin{table}[h]
\centering
\small
\caption{Graph vs raw-text ablation (macro-F1). The SRA graph improves CheXpert-5x200 and CXR-LT by about 11 pp. The gap is smaller on PadChest-GR because its pseudo-reports are sentence-level and rarely contain mixed positive/negative findings.}
\label{tab:rq6}
\begin{tabular}{lccc}
\toprule
Dataset & Raw-text aliases & Radiological Aliases & $\Delta$ macro-F1 \\
\midrule
CheXpert 5x200 & 0.76 & 0.88 & $+$11.7 pp \\
CXR-LT 25-label & 0.50 & 0.61 & $+$11.1 pp \\
PadChest-GR (PadChest synonyms) & 0.62 & 0.62 & $+$0.0 pp \\
\bottomrule
\end{tabular}
\end{table}

The SRA graph improves CheXpert 5x200 and CXR-LT by about 11 F1 points, mostly on labels confused by nearby words: calcification of the aorta, pneumothorax, and lung opacity.
On PadChest-GR the gap disappears because the inputs are already short, single-finding pseudo-reports; there is much less local context for the graph to disambiguate.

\takeawaybox{The SRA graph helps when raw text cannot reliably separate related findings, negations, or anatomy.}

\subsection{Training with extracted labels}
\label{sec:training-results}

\paragraph{Fine labels preserve overall image performance and unlock finer tasks.}
\label{sec:downstream}
We next test whether the extra labels help image models.
We train simple probes on frozen CXR features and compare CheXpert-14, 48 fine leaves, and fine leaves plus parents.
This tests label signal, not leaderboard performance.
Full training details and secondary cross-corpus transfer values are in \Cref{sec:exp-details}.

\begin{table}[h]
\centering
\small
\caption{Image-model summary. Fine labels preserve the overall performance level of the coarse-label baseline while enabling evaluation on fine-grained labels. Detailed probe tables are in \Cref{tab:coarse-vs-fine-full,tab:cxrlt-gold-full}.}
\label{tab:image-summary}
\begin{tabular}{llc}
\toprule
Experiment & Comparison & Macro-AUROC \\
\midrule
HOPPR CXR held-out, DINOv2 & CheXpert-14 vs fine leaves & 0.830 vs \textbf{0.831} \\
HOPPR CXR held-out, DINOv2 & CheXpert-14 vs fine-to-coarse & 0.836 vs 0.834 \\
HOPPR CXR held-out, ConvNeXt-T & CheXpert-14 vs fine leaves & \textbf{0.842} vs 0.826 \\
MIMIC weak $\rightarrow$ CXR-LT gold & ConvNeXt-T fine-label probe & \textbf{0.777} \\
\bottomrule
\end{tabular}
\end{table}

Fine labels do not make image classification harder in the aggregate.
The strict fine-to-coarse test trains on 48 fine labels, aggregates predictions back to the same 13 CheXpert-style targets, and compares against a model trained directly on those 13 targets.
The aggregated fine model reaches 0.834 AUROC versus 0.836 for the direct coarse model.
With DINOv2, the non-aggregated 48-label task also matches the 13-label baseline (0.831 vs 0.830 AUROC). Finally, with ConvNeXt-T, the fine-label model is within 1.6 points of the coarse baseline.
Fine labels do not fragment the signal into a weaker classifier.

\takeawaybox{Fine labels preserve aggregate image-model performance.}

Fine labels also reveal performance CheXpert-14 cannot measure.
The MIMIC-trained ConvNeXt-T probe reaches 0.777 macro-AUROC on expert CXR-LT labels.
Because CXR-LT is a MIMIC subset, this is a gold-label consistency check rather than a leakage-free external validation.
Its best labels are precisely findings CheXpert-14 cannot represent: pneumomediastinum (0.969), subcutaneous emphysema (0.951), pulmonary fibrosis (0.935), pneumoperitoneum (0.885), emphysema (0.873), and calcification of the aorta (0.848).
These are clinically meaningful categories that are invisible in a 14-label benchmark.
The full per-label image table is in \Cref{tab:gold-per-label-full}.

\takeawaybox{Fine labels turn hidden clinical distinctions into measurable image tasks.}

\begin{table}[h]
\centering
\small
\caption{Parent-conditioned child discrimination. Among studies where a parent is positive, the fine-label probe can still distinguish which child finding is present. A model trained only on the parent has no way to do this; chance is 0.5.}
\label{tab:parent-cond}
\begin{tabular}{lrcc}
\toprule
Parent & n(parent+) & \#children & Mean child AUROC \\
\midrule
support\_devices & 586 & 9 & \textbf{0.794 [0.724, 0.861]} \\
pulmonary\_abnormality & 2{,}136 & 12 & 0.735 [0.678, 0.795] \\
cardiomediastinal\_abnormality & 1{,}343 & 6 & 0.708 [0.636, 0.794] \\
airway\_abnormality & 174 & 3 & 0.681 [0.549, 0.760] \\
pleural\_abnormality & 637 & 4 & 0.654 [0.572, 0.735] \\
fracture\_or\_trauma & 274 & 5 & 0.648 [0.501, 0.821] \\
air\_space\_opacity & 696 & 4 & 0.557 [0.509, 0.604] \\
hernia\_abnormality & 101 & 2 & 0.527 [0.161, 0.893] \\
\midrule
\textbf{Mean across parents} & n/a & n/a & \textbf{0.663} \\
\bottomrule
\end{tabular}
\end{table}

This experiment asks the following: given a positive parent, can the model identify the child finding?
This is where fine labels add information.
Within \texttt{support\_devices}, for example, the same parent category contains pacemakers, endotracheal tubes, enteric tubes, central venous catheters, and chest tubes.
A parent-only model can learn ``some support device is present'' but cannot tell which one.
The fine-label probe can assess performance on specific devices. Specifically, support-device children average 0.794 AUROC, with pacemaker 0.964, endotracheal tube 0.913, and enteric tube 0.883.
Similar within-parent structure appears under pulmonary abnormality, where bullous disease reaches 0.940 and pulmonary edema reaches 0.914.

\takeawaybox{Coarse labels indicate a parent is finding positive; fine labels test which child finding the image model can distinguish.}

%% file: tex/7-conclusion.tex
\section{Limitations}
\label{sec:limitations}

\begin{enumerate}[itemsep=2pt, topsep=2pt, leftmargin=*]
\item \textbf{Report-derived, not image-derived.} A study can contain a finding the radiologist did not mention. \emph{Definitely absent} means absent in the report; a missing key means not discussed.
\item \textbf{SRA errors propagate.} If RadGraph-XL misses an entity, status, or relation, aliases cannot recover it. The same applies to any SRA.
\item \textbf{SRAs cost something to build.} Reconfiguring labels is cheap after structured annotations exist, but creating a good SRA requires annotation, validation, prompting, training, or distillation.
\item \textbf{Alias curation is still human work.} Clinicians still decide which phrases count and which phrases should be excluded. Controllability reduces repeated labor; it does not remove clinical judgment.
\item \textbf{Dictionary design choices are pragmatic, not canonical.} Each finding is placed under a single grouping label even when its etiology is broader (for example, \texttt{subcutaneous\_emphysema} sits under \texttt{fracture\_or\_trauma} although it can also be infectious or iatrogenic); some aliases are descriptive radiographic signs rather than definitive diagnoses (for example, ``blunting of the costophrenic angle'' can reflect effusion or pleural thickening); and a few phrases intentionally map to more than one label. These are choices in the released dictionary, editable by design, not fixed properties of the method.
\item \textbf{Evaluation is against benchmark subsets, not a native external set.} Our gold sets are existing benchmarks (CheXpert 5x200, CXR-LT, PadChest-GR); we do not yet have a radiologist-adjudicated external native-report set spanning all 58 labels, which would be required for clinical-utility and generalizability claims. We also report macro-averaged metrics; per-label support, micro-F1, and prevalence-weighted F1, along with an explicit false-positive/error analysis and a documented curation protocol (curator count, freeze date, inter-rater agreement, conflict resolution), are left to a peer-reviewed version.
\item \textbf{Our experiments flatten attributes.} We do not split laterality, severity, acuity, or change over time. This is a choice in the current taxonomy, not a limitation of the method.
\item \textbf{CXR-LT transfer is not fully external.} CXR-LT is a MIMIC-CXR subset. We use it as an expert-label consistency check, not as leakage-free external validation; future image experiments should exclude CXR-LT studies or patients from MIMIC training.
\end{enumerate}

\section{Conclusion}
\label{sec:conclusion}

Radiology reports contain more information than fixed labels expose.
Radiological Aliases turns reusable SRA annotations into local, auditable labels that are cheap to change and broad enough for long-tail findings.
In our experiments, this yields precise report labels in the standard CheXpert setting and competitive precision on long-tail panels, reconfiguration without a repeated full-corpus inference pass, and image models that keep overall performance while learning fine-grained findings.
Once reports are structured, labels no longer have to be scarce, static, or expensive.

%% file: tex/5-appendix.tex
\section{Labeling algorithm: full specification}
\label{sec:algo-appendix}

This appendix states the exact procedure used by Radiological Aliases, including the parts glossed over in \Cref{sec:aliases}.
The implementation is in \texttt{src/create\_weak\_labels.py} (\texttt{label\_study}).

\paragraph{Notation.}
The SRA returns, for a given report, a set of entities.
Each entity has a \texttt{tokens} field, a \texttt{label} of the form \texttt{\{Anatomy|Observation|Observation-Modifier\}::\{definitely present|uncertain|definitely absent\}}, a \texttt{(start\_ix, end\_ix)} token-position range in the report, and a list of outgoing \texttt{relations} of the form \texttt{(relation\_type, target\_entity\_id)} with \texttt{relation\_type} $\in \{\texttt{modify}, \texttt{located\_at}, \texttt{suggestive\_of}, \dots\}$.
All tokens are lowercased and lemmatised (WordNet, default POS, cached) before any matching; the same normalisation is applied to alias phrases.

\paragraph{1. Seeds.}
We iterate over all entities.
An entity is an \emph{observation seed} if (a) its \texttt{label} starts with \texttt{Observation::}, and (b) it has no outgoing \texttt{modify} relation (i.e.\ it is not itself modifying another entity).
Its \emph{status} is read directly from the \texttt{Observation::...} suffix and is one of \texttt{definitely present}, \texttt{uncertain}, \texttt{definitely absent}.
Modifier-only entities (\texttt{Observation-Modifier::...} or observations that only modify another observation, e.g.\ \emph{stable}, \emph{postoperative}) are not seeds; they are still reachable as tokens via relations.

\paragraph{2. Relational neighborhood.}
For each seed we compute a \emph{relational neighborhood}: the set of entities reachable from the seed through any outgoing relation (transitively), together with all entities that have a one-hop \emph{incoming} relation into the seed or into the entities already collected outgoing.
The motivation for the asymmetry is that RadGraph's \texttt{located\_at} edges typically point from observation to anatomy (\emph{effusion} $\xrightarrow{\texttt{located\_at}}$ \emph{pleural}), but modifier anatomies such as a laterality token may appear only as incoming edges.
From the neighborhood we extract two artifacts:
(a) a \texttt{token\_set}, the union of all normalised tokens of all entities in the neighborhood; and
(b) a \texttt{token\_map}, a dictionary from each token to the list of token positions where it appears in the report.
We additionally keep an \texttt{anatomy\_set}, the subset of \texttt{token\_set} coming only from entities with an \texttt{Anatomy::} label, used by the \texttt{require\_anatomy} gating described below.

\paragraph{3. Alias matching.}
For each disease $d$ with alias phrases $\mathcal{A}_d$, and for each seed with token set $T$, the alias $a \in \mathcal{A}_d$ \emph{fires} iff $\text{tokens}(a) \subseteq T$, where $\text{tokens}(a)$ is the normalised-token set of the alias phrase (multiset collapse: each token checked once).
A firing alias records (a) its phrase, (b) the seed's status, and (c) the token positions in the report where its tokens appear (looked up in \texttt{token\_map}; if a token appears multiple times we record the lowest-index occurrence).

\paragraph{4. Precision gates.}
For each disease, we skip a seed that would otherwise fire if either:
(a) any phrase in the disease's \texttt{exclude} list is itself a token-subset of $T$, which means the neighborhood is about a different clinical concept that shares tokens with this alias; or
(b) the disease declares a non-empty \texttt{require\_anatomy} list and none of its tokens appear in the \texttt{anatomy\_set}.
In the frozen taxonomy used in this paper, \texttt{exclude} is set on five disease keys (\texttt{pleural\_effusion}, \texttt{pneumothorax}, \texttt{hilar\_lymphadenopathy}, \texttt{pulmonary\_edema}, \texttt{other\_hernia}) to handle token-overlap collisions identified during development; \texttt{require\_anatomy} is supported by the compiler but left empty in the frozen taxonomy, because anatomy gating on our key labels caused large recall losses during development.

\paragraph{5. Collapse.}
A seed may produce several hits for the same alias; an alias may fire from several seeds.
At each level of aggregation (per alias, then per disease, then per parent) we collapse statuses using the priority \emph{definitely present} $>$ \emph{uncertain} $>$ \emph{definitely absent}.
Diseases with no hit are \emph{omitted} from the output (rather than labeled absent), so a downstream consumer can distinguish ``not discussed in the report'' from ``explicitly stated absent''; treating a missing key as absent would double-count negations.

\paragraph{Output.}
The compiler writes two JSON files per corpus:
one of the shape $\{sid: \{disease: status\}\}$ for the labels, and
one of the shape $\{sid: \{text, [\{disease, alias, label, start\_ix\}, \dots]\}\}$ for the evidence spans.
Every positive label therefore ships with the exact alias phrase and the token positions in the report that produced it.
The alias-set identifier is embedded in each file header.

\paragraph{Uncertainty re-mapping (ablation knob).}
The \texttt{uncertainty\_policy} flag in \texttt{label\_study} optionally rewrites all \emph{uncertain} hits to \emph{present}, \emph{absent}, or drops them entirely before collapse.
The default policy (\emph{keep}) is what we use in all headline results; uncertainty-policy ablations are not part of the main paper.

\clearpage

\section{Alias dictionary summary}
\label{sec:alias-appendix}

The source of truth for the aliases is the Python dictionary in \texttt{src/radlabels/aliases.py}.
Here we show a compact, human-readable summary of the label groups and representative aliases used in the paper.
The table is not exhaustive; it is meant to make the taxonomy inspectable without printing several pages of near-synonyms.
Full label mappings, alias lists, and corpus support counts are shown in \Cref{tab:full-label-mapping,tab:full-alias-dictionary,tab:full-label-support}.

\begin{table}[h]
\centering
\small
\caption{Compact view of the Radiological Aliases dictionary. Parents are derived automatically from the child labels. Examples are representative aliases, not the full dictionary.}
\label{tab:alias-summary}
\begin{tabularx}{\linewidth}{p{0.22\linewidth}p{0.30\linewidth}p{0.38\linewidth}}
\toprule
Parent group & Child labels (examples) & Representative aliases \\
\midrule
Pulmonary abnormality &
\texttt{atelectasis}, \texttt{consolidation}, \texttt{pneumonia}, \texttt{emphysema}, \texttt{pulmonary\_fibrosis}, \texttt{lung\_opacity}, \texttt{lung\_nodule\_or\_mass} &
``atelectasis'', ``lobar collapse'', ``consolidation'', ``pneumonia'', ``emphysema'', ``fibrosis'', ``lung opacity'', ``nodule'', ``mass'' \\
\midrule
Pleural abnormality &
\texttt{pleural\_effusion}, \texttt{pneumothorax}, \texttt{pleural\_thickening}, \texttt{pleural\_other} &
``pleural effusion'', ``pleural fluid'', ``hydrothorax'', ``pneumothorax'', ``pleural thickening'', ``pleural plaque'' \\
\midrule
Cardiomediastinal abnormality &
\texttt{cardiomegaly}, \texttt{enlarged\_cardiomediastinum}, \texttt{tortuous\_aorta}, \texttt{calcification\_of\_the\_aorta}, \texttt{pulmonary\_artery\_enlargement} &
``cardiomegaly'', ``enlarged cardiac silhouette'', ``tortuous aorta'', ``aortic calcification'', ``enlarged pulmonary artery'' \\
\midrule
Airway / hilar abnormality &
\texttt{bronchial\_wall\_thickening}, \texttt{peribronchial\_cuffing}, \texttt{tracheal\_deviation}, \texttt{hilar\_lymphadenopathy} &
``bronchial wall thickening'', ``peribronchial cuffing'', ``tracheal deviation'', ``hilar adenopathy'' \\
\midrule
Fracture or trauma &
\texttt{acute\_rib\_fracture}, \texttt{non\_acute\_rib\_fracture}, \texttt{fracture\_generic}, \texttt{subcutaneous\_emphysema} &
``acute rib fracture'', ``nondisplaced rib fracture'', ``old rib fracture'', ``fracture'', ``subcutaneous emphysema'' \\
\midrule
Mediastinal or abdominal air &
\texttt{pneumomediastinum}, \texttt{pneumoperitoneum} &
``pneumomediastinum'', ``mediastinal air'', ``pneumoperitoneum'', ``free intraperitoneal air'' \\
\midrule
Hernia &
\texttt{hiatus\_hernia}, \texttt{other\_hernia} &
``hiatal hernia'', ``hiatus hernia'', ``diaphragmatic hernia'' \\
\midrule
Support devices &
endotracheal, tracheostomy, enteric and chest tubes; central venous catheters; pacemakers / cardiac device wires &
``endotracheal tube'', ``tracheostomy tube'', ``enteric tube'', ``central venous catheter'', ``pacemaker'', ``chest tube'' \\
\bottomrule
\end{tabularx}
\end{table}

\input{tex/generated_label_support_table}

\clearpage

\section{Experiment details}
\label{sec:exp-details}

\paragraph{LLM cost model.}
Input tokens per report are measured with \texttt{tiktoken cl100k} on 1{,}000 randomly sampled reports from HOPPR CXR; the mean is 162 tokens/report.
For an LLM labeling prompt we assume 800 tokens of instruction/schema overhead and 200 tokens of structured-JSON output per report.
Costs in \Cref{tab:cost} use public vendor prices from 2026-04 and include input plus output tokens.
They are scale estimates, not vendor recommendations.
They assume one report per request and no prompt batching; batching can reduce prompt overhead, but report input/output tokens and recurring API calls still scale with corpus size.

\paragraph{Accuracy metrics.}
For CheXpert 5x200, CXR-LT 2024 Task 2, and PadChest-GR, we report macro precision, recall, and F1 over labels with nonzero support.
Summary-table confidence intervals are 1{,}000-bootstrap 95\% CIs, computed by resampling labels with support $>0$.
Per-label CIs, where reported, are bootstrapped over studies.
Precision is the primary metric because Radiological Aliases is designed to be closed-vocabulary and high-precision: if an alias does not cover a synonym, recall falls, but a fired positive should be easy to audit against the report.

\paragraph{Baselines.}
CheXbert~\citep{smit2020chexbert} is a BERT-based report labeler with a fixed CheXpert-14 output head.
We compare against CheXbert rather than the older CheXpert-NLP rule-based labeler~\citep{irvin2019chexpert}, because CheXbert was trained to dominate CheXpert-NLP on the CheXpert report-labeling benchmark.
Both CheXbert and CheXpert-NLP are structurally confined to CheXpert-style labels, so they score zero on labels they cannot output.
The Raw-text aliases baseline applies the same alias phrases directly to raw report text, without the SRA graph.

\paragraph{Status handling.}
For report-label metrics and image training, only \texttt{definitely present} is treated as positive.
\texttt{uncertain}, \texttt{definitely absent}, and missing labels are treated as negative in the binary matrices, following the common CheXpert U-Zeros convention.
This collapse is used for model training; the raw label files still preserve the distinction between explicitly absent and not detected.

\paragraph{Controllability demo.}
\label{sec:control}
As a lightweight controllability test, we added a new ``LVAD'' (left ventricular assist device) label by writing 12 aliases and recompiling cached annotations.
On 100{,}000 studies, the recompile took 85 seconds and found 7 positive LVAD cases with evidence spans.
Relabeling the same 100{,}000 reports with Claude Opus would cost \$2{,}943 and take hours.
We keep this result in the appendix because the main text already motivates controllability; the central experiments focus on label quality and image-model utility.

\paragraph{Image probe training.}
We extract frozen image features from DINOv2-ViT-B/14~\citep{oquab2024dinov2} and ConvNeXt-Tiny~\citep{liu2022convnext}.
For within-corpus experiments, studies are split deterministically by study ID into 80\% train, 10\% validation, and 10\% test.
The probe is a two-layer MLP (feature dimension $\rightarrow$ 512 $\rightarrow$ number of labels), trained with BCEWithLogitsLoss using per-class positive weights (negative/positive ratio, clipped at 100).
We use AdamW, train for up to 15 epochs, and early-stop on validation macro-AUROC.
Bootstrap confidence intervals for image experiments are computed over labels with support $>0$.

\paragraph{Fine-to-coarse aggregation.}
The aggregation experiment is implemented in \texttt{src/fine\_to\_coarse\_aggregation.py}.
It trains one probe on 48 fine labels and one probe directly on the 13 CheXpert-style targets, using the same HOPPR CXR split.
Fine predictions are mapped to the CheXpert-style target space by using the leaf probability directly when the target is a leaf, and max child probability when the target is a parent.

\paragraph{Cross-corpus weak-label transfer.}
In addition to the headline CXR-LT gold transfer in \Cref{tab:image-summary}, we trained the same DINOv2 probe on one weak-label corpus and evaluated on another.
HOPPR CXR $\rightarrow$ MIMIC-CXR yields AUROC 0.729 / AUPRC 0.127 over 48 leaves; MIMIC-CXR $\rightarrow$ HOPPR CXR yields AUROC 0.777 / AUPRC 0.121; MIMIC-CXR $\rightarrow$ CheXpert+ yields AUROC 0.721 / AUPRC 0.139.
The drop from within-corpus to cross-corpus evaluation is expected, but no transfer collapses.

\begin{table}[h]
\centering
\small
\caption{Coarse versus fine labels on the HOPPR CXR held-out split. Fine labels match CheXpert-14 AUROC while exposing many more output heads.}
\label{tab:coarse-vs-fine-full}
\begin{tabular}{lrcc}
\toprule
Target label set & \#heads & DINOv2 AUROC & ConvNeXt-T AUROC \\
\midrule
CheXpert-14 & 13 & 0.830 [0.789, 0.872] & \textbf{0.842 [0.800, 0.883]} \\
Fine leaves & 48 & \textbf{0.831 [0.805, 0.857]} & 0.826 [0.789, 0.857] \\
Fine + parents & 58 & 0.824 [0.801, 0.847] & n/a \\
\bottomrule
\end{tabular}
\end{table}

\begin{table}[h]
\centering
\small
\caption{Weak-label train $\rightarrow$ gold-label test on CXR-LT 2024 Task 2. Probes trained on our weak labels transfer to radiologist-adjudicated labels, especially when trained on MIMIC-CXR from the same source institution.}
\label{tab:cxrlt-gold-full}
\begin{tabular}{llcc}
\toprule
Train corpus & Backbone & Test AUROC & Test AUPRC \\
\midrule
\textbf{MIMIC-CXR (212k)} & DINOv2 & 0.729 [0.683, 0.770] & 0.340 [0.247, 0.428] \\
\textbf{MIMIC-CXR (212k)} & ConvNeXt-T & \textbf{0.777 [0.734, 0.816]} & \textbf{0.371 [0.282, 0.461]} \\
HOPPR CXR (86k) & DINOv2 & 0.655 [0.620, 0.692] & 0.285 [0.205, 0.365] \\
HOPPR CXR (86k) & ConvNeXt-T & 0.689 [0.648, 0.732] & 0.299 [0.218, 0.384] \\
CheXpert+ (187k) & DINOv2 & 0.663 [0.609, 0.711] & 0.290 [0.208, 0.373] \\
CheXpert+ (187k) & ConvNeXt-T & 0.691 [0.642, 0.737] & 0.331 [0.250, 0.419] \\
\bottomrule
\end{tabular}
\end{table}

\clearpage

\section{Full per-label tables}

\begin{table}[h]
\centering
\small
\caption{CXR-LT 2024 Task 2, all 25 labels. Per-label precision/recall/F1 for Radiological Aliases, plus CheXbert F1 and Raw-text aliases F1 for comparison. CheXbert $F_1 = 0$ on 12 labels (structurally absent from its 14-output head).}
\label{tab:cxrlt-full}
\begin{tabular}{lrrrrrr}
\toprule
Label & Support & Ours P & Ours R & Ours F1 & CheXbert F1 & Raw-text F1 \\
\midrule
Atelectasis & 125 & 0.52 & 0.67 & 0.59 & 0.63 & 0.59 \\
Calcification of the Aorta & 47 & 0.94 & 0.66 & 0.78 & 0.00 & 0.19 \\
Cardiomegaly & 159 & 0.88 & 0.64 & 0.74 & 0.76 & 0.65 \\
Consolidation & 74 & 0.87 & 0.73 & 0.79 & 0.78 & 0.57 \\
Edema & 101 & 0.73 & 0.47 & 0.57 & 0.80 & 0.43 \\
Emphysema & 29 & 0.36 & 0.76 & 0.49 & 0.00 & 0.49 \\
Enlarged Cardiomediastinum & 118 & 1.00 & 0.03 & 0.07 & 0.60 & 0.00 \\
Fibrosis & 22 & 1.00 & 0.64 & 0.78 & 0.00 & 0.86 \\
Fracture & 48 & 0.53 & 0.85 & 0.65 & 0.89 & 0.64 \\
Hernia & 19 & 0.88 & 0.74 & 0.80 & 0.00 & 0.83 \\
Infiltration & 12 & 0.50 & 0.58 & 0.54 & 0.00 & 0.44 \\
Lung Lesion & 6 & 1.00 & 0.33 & 0.50 & 0.10 & 0.29 \\
Lung Opacity & 199 & 0.93 & 0.35 & 0.51 & 0.84 & 0.18 \\
Mass & 20 & 0.21 & 0.35 & 0.26 & 0.00 & 0.19 \\
Nodule & 33 & 0.79 & 0.79 & 0.79 & 0.00 & 0.75 \\
Pleural Effusion & 183 & 0.85 & 0.81 & 0.83 & 0.83 & 0.57 \\
Pleural Other & 19 & 0.00 & 0.00 & 0.00 & 0.44 & 0.00 \\
Pleural Thickening & 22 & 0.61 & 0.91 & 0.73 & 0.00 & 0.78 \\
Pneumomediastinum & 35 & 0.90 & 0.80 & 0.85 & 0.00 & 0.82 \\
Pneumonia & 22 & 0.17 & 0.64 & 0.26 & 0.24 & 0.21 \\
Pneumoperitoneum & 24 & 0.89 & 0.67 & 0.76 & 0.00 & 0.75 \\
Pneumothorax & 49 & 0.85 & 0.69 & 0.76 & 0.79 & 0.26 \\
Subcutaneous Emphysema & 42 & 0.85 & 0.83 & 0.84 & 0.00 & 0.89 \\
Support Devices & 221 & 0.98 & 0.65 & 0.78 & 0.88 & 0.81 \\
Tortuous Aorta & 33 & 0.81 & 0.39 & 0.53 & 0.00 & 0.25 \\
\bottomrule
\end{tabular}
\end{table}

\clearpage

\begin{table}[h]
\centering
\small
\caption{CXR-LT gold image classification, all 21 evaluated labels. Test AUROC/AUPRC of a ConvNeXt-T probe trained on MIMIC-CXR weak labels. Sorted by AUROC.}
\label{tab:gold-per-label-full}
\begin{tabular}{lrrr}
\toprule
Our taxonomy key & Support & AUROC & AUPRC \\
\midrule
pneumomediastinum & 24 & 0.969 & 0.631 \\
subcutaneous\_emphysema & 30 & 0.951 & 0.678 \\
pulmonary\_fibrosis & 12 & 0.935 & 0.402 \\
pneumoperitoneum & 24 & 0.885 & 0.375 \\
emphysema & 24 & 0.873 & 0.232 \\
pneumothorax & 34 & 0.861 & 0.579 \\
calcification\_of\_the\_aorta & 39 & 0.848 & 0.496 \\
pulmonary\_edema & 96 & 0.812 & 0.551 \\
tortuous\_aorta & 36 & 0.789 & 0.311 \\
pleural\_effusion & 153 & 0.784 & 0.685 \\
pleural\_thickening & 22 & 0.775 & 0.161 \\
consolidation & 60 & 0.772 & 0.365 \\
cardiomegaly & 123 & 0.759 & 0.576 \\
lung\_nodule\_or\_mass & 35 & 0.746 & 0.228 \\
lung\_lesion & 4 & 0.732 & 0.040 \\
infiltration & 12 & 0.719 & 0.070 \\
pneumonia & 17 & 0.693 & 0.090 \\
atelectasis & 103 & 0.663 & 0.400 \\
lung\_opacity & 165 & 0.658 & 0.585 \\
pleural\_other & 12 & 0.579 & 0.060 \\
enlarged\_cardiomediastinum & 99 & 0.519 & 0.281 \\
\bottomrule
\end{tabular}
\end{table}

%% file: tex/generated_label_support_table.tex
\begingroup
\tiny
\setlength{\tabcolsep}{1.5pt}
\setlength{\LTleft}{0pt}
\setlength{\LTright}{0pt}
\setlength{\LTcapwidth}{\linewidth}
\refstepcounter{table}\label{tab:full-label-mapping}
\noindent\textbf{Table~\thetable. Full label mapping.} For each canonical label, we report its parent label, the number of alias phrases in the dictionary, and whether it maps to CheXpert-14 and/or CXR-LT.
\begin{longtable}{@{}p{0.24\linewidth}p{0.18\linewidth}p{0.08\linewidth}p{0.21\linewidth}p{0.23\linewidth}@{}}
\toprule
Label & Parent & \# aliases & CheXpert map & CXR-LT map \\
\midrule
\endfirsthead
\toprule
Label & Parent & \# aliases & CheXpert map & CXR-LT map \\
\midrule
\endhead
\bottomrule
\endfoot
\path{acute_rib_fracture} & \path{fracture_or_trauma} & 8 &  &  \\
\path{air_space_opacity} & \path{pulmonary_abnormality} & 32 &  &  \\
\path{airway_abnormality} &  & 0 &  &  \\
\path{atelectasis} & \path{pulmonary_abnormality} & 21 & Atelectasis & Atelectasis \\
\path{bronchial_wall_thickening} & \path{airway_abnormality} & 7 &  &  \\
\path{bullous_disease} & \path{pulmonary_abnormality} & 13 &  &  \\
\path{calcification_of_the_aorta} & \path{cardiomediastinal_abnormality} & 6 &  & Calcification of the Aorta \\
\path{cardiomediastinal_abnormality} &  & 0 &  &  \\
\path{cardiomegaly} & \path{cardiomediastinal_abnormality} & 18 & Cardiomegaly & Cardiomegaly \\
\path{central_venous_catheter} & \path{support_devices} & 17 &  &  \\
\path{consolidation} & \path{air_space_opacity} & 7 & Consolidation & Consolidation \\
\path{emphysema} & \path{pulmonary_abnormality} & 6 &  & Emphysema \\
\path{endotracheal_tube} & \path{support_devices} & 6 &  &  \\
\path{enlarged_cardiomediastinum} & \path{cardiomediastinal_abnormality} & 7 & Enlarged Cardiomediastinum & Enlarged Cardiomediastinum \\
\path{enteric_tube} & \path{support_devices} & 8 &  &  \\
\path{fracture_generic} & \path{fracture_or_trauma} & 11 &  &  \\
\path{fracture_or_trauma} &  & 0 & Fracture & Fracture \\
\path{ground_glass_opacity} & \path{air_space_opacity} & 12 &  &  \\
\path{hernia_abnormality} &  & 0 &  & Hernia \\
\path{hiatus_hernia} & \path{hernia_abnormality} & 10 &  &  \\
\path{hilar_abnormality} &  & 0 &  &  \\
\path{hilar_lymphadenopathy} & \path{hilar_abnormality} & 14 &  &  \\
\path{hyperinflation} & \path{pulmonary_abnormality} & 20 &  &  \\
\path{implantable_electronic_device} & \path{support_devices} & 14 &  &  \\
\path{infiltration} & \path{air_space_opacity} & 7 &  & Infiltration \\
\path{intercostal_drain} & \path{support_devices} & 16 &  &  \\
\path{interstitial_thickening} & \path{pulmonary_abnormality} & 19 &  &  \\
\path{lobar_segmental_collapse} & \path{pulmonary_abnormality} & 32 &  &  \\
\path{lung_lesion} & \path{pulmonary_abnormality} & 6 & Lung Lesion & Lung Lesion \\
\path{lung_nodule_or_mass} & \path{pulmonary_abnormality} & 21 &  & Mass, Nodule \\
\path{lung_opacity} & \path{pulmonary_abnormality} & 8 & Lung Opacity & Lung Opacity \\
\path{mediastinal_or_abdominal_air} &  & 0 &  &  \\
\path{non_acute_rib_fracture} & \path{fracture_or_trauma} & 15 &  &  \\
\path{nonsurgical_internal_foreign_body} & \path{support_devices} & 29 &  &  \\
\path{other_hernia} & \path{hernia_abnormality} & 4 &  &  \\
\path{pacemaker_electronic_cardiac_device_or_wires} & \path{support_devices} & 26 &  &  \\
\path{peribronchial_cuffing} & \path{airway_abnormality} & 6 &  &  \\
\path{pleural_abnormality} &  & 0 &  &  \\
\path{pleural_effusion} & \path{pleural_abnormality} & 24 & Pleural Effusion & Pleural Effusion \\
\path{pleural_other} & \path{pleural_abnormality} & 6 & Pleural Other & Pleural Other \\
\path{pleural_thickening} & \path{pleural_abnormality} & 7 &  & Pleural Thickening \\
\path{pneumomediastinum} & \path{mediastinal_or_abdominal_air} & 5 &  & Pneumomediastinum \\
\path{pneumonia} & \path{air_space_opacity} & 10 & Pneumonia & Pneumonia \\
\path{pneumoperitoneum} & \path{mediastinal_or_abdominal_air} & 6 &  & Pneumoperitoneum \\
\path{pneumothorax} & \path{pleural_abnormality} & 12 & Pneumothorax & Pneumothorax \\
\path{pulmonary_abnormality} &  & 0 &  &  \\
\path{pulmonary_artery_enlargement} & \path{cardiomediastinal_abnormality} & 20 &  &  \\
\path{pulmonary_congestion_pulmonary_venous_congestion} & \path{cardiomediastinal_abnormality} & 19 &  &  \\
\path{pulmonary_edema} & \path{pulmonary_abnormality} & 8 & Edema & Edema \\
\path{pulmonary_fibrosis} & \path{pulmonary_abnormality} & 15 &  & Fibrosis \\
\path{shoulder_dislocation} & \path{fracture_or_trauma} & 19 &  &  \\
\path{subcutaneous_emphysema} & \path{fracture_or_trauma} & 16 &  & Subcutaneous Emphysema \\
\path{support_devices} &  & 0 & Support Devices & Support Devices \\
\path{support_devices_generic} & \path{support_devices} & 6 &  &  \\
\path{tortuous_aorta} & \path{cardiomediastinal_abnormality} & 10 &  & Tortuous Aorta \\
\path{tracheal_deviation} & \path{airway_abnormality} & 15 &  &  \\
\path{tracheostomy_tube} & \path{support_devices} & 4 &  &  \\
\path{whole_lung_or_majority_collapse} & \path{pulmonary_abnormality} & 21 &  &  \\
\end{longtable}
\clearpage
\refstepcounter{table}\label{tab:full-alias-dictionary}
\noindent\textbf{Table~\thetable. Full alias dictionary.} Each row shows the exact alias phrases used for one canonical label. Parent labels have no direct aliases and are omitted.
\begin{longtable}{@{}p{0.24\linewidth}p{0.72\linewidth}@{}}
\toprule
Label & Alias phrases \\
\midrule
\endfirsthead
\toprule
Label & Alias phrases \\
\midrule
\endhead
\bottomrule
\endfoot
\path{acute_rib_fracture} & acute rib fracture; new rib fracture; recent rib fracture; displaced rib fracture; nondisplaced rib fracture; non-displaced rib fracture; minimally displaced rib fracture; acute rib fractures \\
\path{air_space_opacity} & consolidation; infiltrate; pneumonia; airspace disease; airspace opacity; air space disease; air space opacity; air-space disease; air-space opacity; groundglass opacity; ground-glass opacity; ground glass opacity; ground glass opacities; ground-glass opacities; patchy opacity; patchy opacities; confluent opacity; confluent opacities; alveolar opacity; alveolar opacities; alveolar infiltrate; parenchymal opacity; airspace infiltrate; airspace process; parenchymal infiltrate; focal consolidation; lobar consolidation; patchy consolidation; dense consolidation; pneumonic infiltrate; pneumonic consolidation; bronchopneumonia \\
\path{atelectasis} & atelectasis; atelectatic; atelectatic lung; lung collapse; collapsed lung; linear atelectasis; discoid atelectasis; plate atelectasis; plate-like atelectasis; platelike atelectasis; bandlike atelectasis; band-like atelectasis; streaky atelectasis; bibasilar atelectasis; basilar atelectasis; bilateral atelectasis; scarring atelectasis; compressive atelectasis; relaxation atelectasis; passive atelectasis; rounded atelectasis \\
\path{bronchial_wall_thickening} & bronchial wall thickening; bronchial thickening; airway wall thickening; airway thickening; thickened bronchial walls; thickened airway walls; bronchial wall thickened \\
\path{bullous_disease} & bullous changes; bullous change; pulmonary bullae; pulmonary bulla; bullae; bullous emphysema; giant bulla; giant bullae; bullous lung disease; apical bullae; apical bulla; subpleural bullae; subpleural bulla \\
\path{calcification_of_the_aorta} & aortic calcification; calcified aorta; atherosclerotic calcification; atherosclerotic aortic calcification; aortic atherosclerosis; aortic knob calcification \\
\path{cardiomegaly} & cardiomegaly; enlarged heart; heart enlarged; cardiac enlargement; cardiac silhouette enlargement; heart enlargement; increased cardiac; increased heart size; enlarged cardiac silhouette; enlarged cardiomediastinal silhouette; dilated heart; cardiac hypertrophy; globular heart; large heart; large cardiac silhouette; cardiac silhouette enlarged; heart size enlarged; prominent cardiac silhouette \\
\path{central_venous_catheter} & central venous catheter; central line; cvc; picc line; picc; peripherally inserted central catheter; subclavian line; internal jugular line; tunneled catheter; hickman catheter; port-a-cath; portacath; implanted port; mediport; triple lumen catheter; swan-ganz catheter; pulmonary artery catheter \\
\path{consolidation} & consolidation; focal consolidation; lobar consolidation; patchy consolidation; dense consolidation; pneumonic consolidation; consolidative opacity \\
\path{emphysema} & emphysema; centrilobular emphysema; paraseptal emphysema; panlobular emphysema; emphysematous changes; copd emphysema \\
\path{endotracheal_tube} & endotracheal tube; et tube; ett; endotracheal intubation; tracheal tube; endotracheal cannula \\
\path{enlarged_cardiomediastinum} & enlarged cardiomediastinum; enlarged cardiomediastinal silhouette; widened mediastinum; mediastinal widening; mediastinal enlargement; prominent mediastinum; prominent mediastinal silhouette \\
\path{enteric_tube} & enteric tube; nasogastric tube; ng tube; orogastric tube; og tube; dobhoff tube; feeding tube; gastric tube \\
\path{fracture_generic} & fracture; fractures; osseous fracture; clavicle fracture; clavicular fracture; humerus fracture; humeral fracture; scapular fracture; sternal fracture; vertebral fracture; compression fracture \\
\path{ground_glass_opacity} & ground glass opacity; ground-glass opacity; groundglass opacity; ground glass opacities; ground-glass opacities; ground glass pattern; ground glass infiltrate; ground glass infiltrates; ground glass nodule; ground glass; hazy opacity; hazy opacities \\
\path{hiatus_hernia} & hiatal hernia; hiatus hernia; paraesophageal hernia; para-esophageal hernia; gastric hernia; sliding hiatal hernia; large hiatal hernia; small hiatal hernia; retrocardiac hiatal hernia; hernia hiatus \\
\path{hilar_lymphadenopathy} & hilar adenopathy; hilar lymphadenopathy; enlarged hilar nodes; enlarged hilar lymph nodes; hilar lymph node enlargement; prominent hilar nodes; prominent hilar lymph nodes; bulky hilar nodes; bulky hila; bulky hilum; lymphadenopathy hilum; adenopathy hilum; hilar mass; bilateral hilar adenopathy \\
\path{hyperinflation} & hyperexpanded; hyperexpansion; hyperinflated; hyperinflation; increased lung volume; increased lung volumes; overinflated lungs; overinflation; overexpanded; hyperaeration; hyperaerated; barrel chest; flattened diaphragm; flattened diaphragms; flattened hemidiaphragm; flattened hemidiaphragms; flattening diaphragm; flattening hemidiaphragms; low flat diaphragms; lung hyperexpansion \\
\path{implantable_electronic_device} & spinal cord stimulator; neurostimulator; nerve stimulator; pain pump; medicine pump; intrathecal pump; infusion pump; loop recorder; implantable loop recorder; dorsal column stimulator; vagal nerve stimulator; vagus nerve stimulator; deep brain stimulator; phrenic nerve stimulator \\
\path{infiltration} & infiltrate; infiltrates; pulmonary infiltrate; alveolar infiltrate; interstitial infiltrate; airspace infiltrate; parenchymal infiltrate \\
\path{intercostal_drain} & pleural drain; pleural tube; chest tube; chest tubes; chest drain; chest drains; intercostal tube; intercostal drain; thoracostomy tube; thoracotomy tube; thoracotomy drain; thracostomy drain; pigtail catheter; pigtail drain; pleurx catheter; tunneled pleural catheter \\
\path{interstitial_thickening} & interstitial markings; reticular markings; reticular pattern; reticulonodular pattern; reticulonodular markings; interstitial pattern; interstitial changes; septal thickening; septal lines; kerley lines; kerley b lines; thickened interstitium; thickening interstitium; prominent interstitium; interstitial prominence; interstitial prominence lungs; crazy paving; honeycombing; fibrosis interstitial \\
\path{lobar_segmental_collapse} & lobar atelectasis; lobar collapse; segmental atelectasis; segmental collapse; subsegmental atelectasis; subsegmental collapse; segment atelectasis; atelectasis segment; right lower lobe atelectasis; right middle lobe atelectasis; right upper lobe atelectasis; left upper lobe atelectasis; left lower lobe atelectasis; right lower lobe collapse; right middle lobe collapse; right upper lobe collapse; left upper lobe collapse; left lower lobe collapse; lingular atelectasis; lingula atelectasis; lingular collapse; lingula collapse; rll atelectasis; rml atelectasis; rul atelectasis; lll atelectasis; lul atelectasis; rll collapse; rml collapse; rul collapse; lll collapse; lul collapse \\
\path{lung_lesion} & lung lesion; pulmonary lesion; parenchymal lesion; cavitary lesion; lytic lesion; cavitation \\
\path{lung_nodule_or_mass} & lung nodule; lung nodules; pulmonary nodule; pulmonary nodules; lung mass; pulmonary mass; parenchymal mass; parenchymal nodule; nodular opacity; nodular density; granuloma; calcified granuloma; calcified nodule; noncalcified nodule; solitary pulmonary nodule; solitary nodule; lung lesion; pulmonary lesion; lung tumor; pulmonary tumor; coin lesion \\
\path{lung_opacity} & lung opacity; lung opacities; pulmonary opacity; pulmonary opacities; parenchymal opacity; focal opacity; diffuse opacity; patchy opacity \\
\path{non_acute_rib_fracture} & healed rib fracture; healed rib fractures; old rib fracture; old rib fractures; chronic rib fracture; chronic rib fractures; subacute rib fracture; healing rib fracture; distant rib fracture; rib fracture callus; callus rib fracture; remote rib fracture; remote rib fractures; prior rib fracture; prior rib fractures \\
\path{nonsurgical_internal_foreign_body} & foreign body; foreign bodies; ingested foreign body; aspirated foreign body; retained foreign body; coin; swallowed coin; ingested object; swallowed object; retained bullet; retained bullets; retained metal fragment; metallic foreign body; metallic fragment; shrapnel; bb pellet; bb pellets; bullet fragment; bullet fragments; fish bone; chicken bone; animal bone; quarter; penny; dime; nickel; button battery; retained BB; retained BBs \\
\path{other_hernia} & diaphragmatic hernia; bochdalek hernia; morgagni hernia; abdominal hernia \\
\path{pacemaker_electronic_cardiac_device_or_wires} & pacemaker; dual chamber pacemaker; single chamber pacemaker; biventricular pacemaker; AICD; CRT; CRT-D; CRT-P; ICD; pacer; defibrillator; implantable defibrillator; implantable cardioverter defibrillator; cardioverter defibrillator; pacing leads; pacing lead; pacemaker leads; pacemaker lead; pacing wires; pacing wire; pacemaker wires; pacemaker wire; epicardial pacing wires; epicardial pacing wire; transvenous pacing lead; leadless pacemaker \\
\path{peribronchial_cuffing} & peribronchial cuffing; peribronchial thickening; peribronchial markings; peribronchial prominence; bronchial cuffing; donut sign \\
\path{pleural_effusion} & pleural effusion; pleural fluid; pleural collection; hemothorax; empyema; hydrothorax; fluid pleural space; loculated effusion; bilateral effusions; layering effusion; free flowing effusion; free-flowing effusion; simple effusion; complex effusion; small effusion; moderate effusion; large effusion; trace effusion; tiny effusion; blunting costophrenic angle; costophrenic blunting; blunted costophrenic angle; layering pleural; meniscus sign \\
\path{pleural_other} & pleural abnormality; pleural disease; pleural lesion; pleural mass; pleural calcification; calcified pleura \\
\path{pleural_thickening} & pleural thickening; thickened pleura; pleural plaque; pleural plaques; calcified pleural plaque; apical pleural thickening; pleural scarring \\
\path{pneumomediastinum} & pneumomediastinum; mediastinal air; mediastinal emphysema; air in the mediastinum; pneumopericardium \\
\path{pneumonia} & pneumonia; bronchopneumonia; atypical pneumonia; lobar pneumonia; bacterial pneumonia; viral pneumonia; aspiration pneumonia; infectious process; infection; pneumonic infiltrate \\
\path{pneumoperitoneum} & pneumoperitoneum; free intraperitoneal air; free abdominal air; free air under the diaphragm; subdiaphragmatic free air; intraperitoneal free air \\
\path{pneumothorax} & pneumothorax; ptx; air pleural space; pleural air; hydropneumothorax; tension pneumothorax; apical pneumothorax; small pneumothorax; large pneumothorax; residual pneumothorax; loculated pneumothorax; basilar pneumothorax \\
\path{pulmonary_artery_enlargement} & enlarged pulmonary arteries; pulmonary artery dilation; pulmonary artery dilatation; prominent hilar vessels; pulmonary artery enlargement; dilated pulmonary artery; dilated pulmonary arteries; enlarged pulmonary artery; prominent pulmonary arteries; prominent pulmonary artery; pulmonary arterial enlargement; hilar vessel prominence; enlarged hilar vessels; pulmonary trunk dilation; pulmonary trunk dilatation; pulmonary trunk enlargement; pulmonary trunk prominence; main pulmonary artery enlargement; main pulmonary artery prominence; central pulmonary artery enlargement \\
\path{pulmonary_congestion_pulmonary_venous_congestion} & congestion; pulmonary congestion; pulmonary venous congestion; pvc; interstitial edema; interstitial pulmonary edema; pulmonary edema; pulmonary interstitial edema; perihilar edema; vascular redistribution; vascular engorgement; vascular congestion; cephalization; cephalization pulmonary vasculature; upper zone redistribution; upper lobe redistribution; prominent pulmonary vasculature; increased pulmonary vasculature; cardiogenic pulmonary edema \\
\path{pulmonary_edema} & pulmonary edema; pulmonary oedema; interstitial pulmonary edema; alveolar pulmonary edema; cardiogenic pulmonary edema; flash pulmonary edema; acute pulmonary edema; fluid overload \\
\path{pulmonary_fibrosis} & pulmonary fibrosis; lung fibrosis; fibrotic change; fibrotic changes; fibrosis; fibrotic scarring; fibrotic band; fibrotic bands; pleuroparenchymal tract; pleuroparenchymal tracts; pleuroparenchymal changes; pleuroparenchymal change; honeycombing; honeycomb lung; usual interstitial pneumonia \\
\path{shoulder_dislocation} & glenohumeral joint dislocation; glenohumeral dislocation; shoulder dislocation; shoulder is dislocated; shoulder dislocated; shoulder subluxation; glenohumeral subluxation; dislocation shoulder; dislocation glenohumeral; humeral head dislocation; humerus dislocation; humeral head subluxed; humeral head subluxation; anterior shoulder dislocation; posterior shoulder dislocation; inferior shoulder dislocation; anteriorly dislocated shoulder; posteriorly dislocated shoulder; dislocated humeral head \\
\path{subcutaneous_emphysema} & subcutaneous emphysema; subcutaneous gas; subcutaneous air; soft tissue emphysema; soft tissue gas; soft tissue air; gas soft tissues; emphysema soft tissues; chest wall emphysema; chest wall gas; chest wall air; subq emphysema; subq air; extensive subcutaneous emphysema; extensive subcutaneous air; subcutaneous emphysema chest wall \\
\path{support_devices_generic} & support device; support devices; hardware; monitoring device; lines and tubes; tubes and lines \\
\path{tortuous_aorta} & tortuous aorta; aortic tortuosity; ectatic aorta; aortic ectasia; aortic elongation; elongated aorta; aorta is elongated; unfolded aorta; unfolding of the aorta; unfolding aorta \\
\path{tracheal_deviation} & tracheal deviation; trachea deviated; deviated trachea; shifted trachea; trachea shifted; leftward trachea; rightward trachea; leftward tracheal deviation; rightward tracheal deviation; displaced trachea; trachea displaced; mass effect trachea; tracheal shift; shift trachea; midline shift trachea \\
\path{tracheostomy_tube} & tracheostomy tube; tracheostomy; trach tube; trach \\
\path{whole_lung_or_majority_collapse} & complete right side atelectasis; total right lung collapse; complete left side atelectasis; total left lung collapse; complete lung collapse; complete lung atelectasis; total lung collapse; total lung atelectasis; whole lung collapse; whole lung atelectasis; entire lung collapse; entire lung atelectasis; complete right lung collapse; complete left lung collapse; whole right lung collapse; whole left lung collapse; entire right lung collapse; entire left lung collapse; collapsed lung; collapse entire lung; collapse whole lung \\
\end{longtable}
\refstepcounter{table}\label{tab:full-label-support}
\noindent\textbf{Table~\thetable. Positive support counts by corpus.} Counts are study-level positives for each canonical label.
\begin{longtable}{@{}p{0.48\linewidth}rrrr@{}}
\toprule
Label & HOPPR & MIMIC & CheX+ & PadChest \\
\midrule
\endfirsthead
\toprule
Label & HOPPR & MIMIC & CheX+ & PadChest \\
\midrule
\endhead
\bottomrule
\endfoot
\path{acute_rib_fracture} & 1,184 & 788 & 677 & 0 \\
\path{air_space_opacity} & 44,462 & 46,092 & 58,026 & 333 \\
\path{airway_abnormality} & 10,761 & 1,655 & 1,197 & 0 \\
\path{atelectasis} & 34,283 & 68,287 & 57,018 & 237 \\
\path{bronchial_wall_thickening} & 2,146 & 373 & 219 & 0 \\
\path{bullous_disease} & 693 & 324 & 205 & 0 \\
\path{calcification_of_the_aorta} & 34,442 & 6,913 & 3,046 & 242 \\
\path{cardiomediastinal_abnormality} & 98,777 & 77,482 & 73,211 & 1,089 \\
\path{cardiomegaly} & 47,704 & 47,905 & 29,620 & 498 \\
\path{central_venous_catheter} & 5,547 & 16,277 & 29,630 & 77 \\
\path{consolidation} & 6,298 & 14,546 & 29,438 & 41 \\
\path{emphysema} & 5,701 & 5,404 & 5,756 & 0 \\
\path{endotracheal_tube} & 5,009 & 20,999 & 19,396 & 39 \\
\path{enlarged_cardiomediastinum} & 1,033 & 2,177 & 2,485 & 25 \\
\path{enteric_tube} & 4,497 & 23,312 & 25,557 & 53 \\
\path{fracture_generic} & 14,707 & 7,430 & 7,999 & 46 \\
\path{fracture_or_trauma} & 15,441 & 10,597 & 13,441 & 151 \\
\path{ground_glass_opacity} & 1,494 & 1,658 & 2,614 & 11 \\
\path{hernia_abnormality} & 5,954 & 2,055 & 929 & 86 \\
\path{hiatus_hernia} & 5,743 & 1,878 & 902 & 86 \\
\path{hilar_abnormality} & 1,016 & 1,143 & 478 & 77 \\
\path{hilar_lymphadenopathy} & 1,016 & 1,143 & 478 & 77 \\
\path{hyperinflation} & 38,741 & 9,822 & 1,565 & 246 \\
\path{implantable_electronic_device} & 524 & 132 & 63 & 4 \\
\path{infiltration} & 21,797 & 2,061 & 1,287 & 152 \\
\path{intercostal_drain} & 2,848 & 12,340 & 22,848 & 7 \\
\path{interstitial_thickening} & 21,914 & 4,642 & 11,452 & 206 \\
\path{lobar_segmental_collapse} & 5,907 & 2,192 & 1,862 & 15 \\
\path{lung_lesion} & 886 & 781 & 363 & 0 \\
\path{lung_nodule_or_mass} & 24,987 & 7,687 & 6,412 & 237 \\
\path{lung_opacity} & 11,537 & 27,957 & 28,673 & 109 \\
\path{mediastinal_or_abdominal_air} & 229 & 1,444 & 1,387 & 0 \\
\path{non_acute_rib_fracture} & 6,094 & 3,056 & 2,014 & 107 \\
\path{nonsurgical_internal_foreign_body} & 767 & 162 & 301 & 0 \\
\path{other_hernia} & 233 & 187 & 29 & 0 \\
\path{pacemaker_electronic_cardiac_device_or_wires} & 14,083 & 12,211 & 14,036 & 109 \\
\path{peribronchial_cuffing} & 8,035 & 580 & 681 & 0 \\
\path{pleural_abnormality} & 38,318 & 67,627 & 93,200 & 581 \\
\path{pleural_effusion} & 22,907 & 59,183 & 81,244 & 370 \\
\path{pleural_other} & 1,704 & 998 & 712 & 0 \\
\path{pleural_thickening} & 15,148 & 4,729 & 4,108 & 228 \\
\path{pneumomediastinum} & 137 & 758 & 852 & 0 \\
\path{pneumonia} & 18,500 & 27,465 & 16,656 & 0 \\
\path{pneumoperitoneum} & 95 & 722 & 568 & 0 \\
\path{pneumothorax} & 2,943 & 9,160 & 17,245 & 11 \\
\path{pulmonary_abnormality} & 143,201 & 123,461 & 125,426 & 1,101 \\
\path{pulmonary_artery_enlargement} & 143 & 191 & 168 & 0 \\
\path{pulmonary_congestion_pulmonary_venous_congestion} & 14,008 & 40,191 & 53,439 & 208 \\
\path{pulmonary_edema} & 3,125 & 26,149 & 47,997 & 0 \\
\path{pulmonary_fibrosis} & 7,468 & 1,276 & 1,000 & 153 \\
\path{shoulder_dislocation} & 100 & 91 & 114 & 0 \\
\path{subcutaneous_emphysema} & 730 & 3,414 & 5,835 & 3 \\
\path{support_devices} & 31,137 & 70,741 & 84,225 & 255 \\
\path{support_devices_generic} & 3,339 & 9,739 & 6,646 & 0 \\
\path{tortuous_aorta} & 27,128 & 8,995 & 3,630 & 465 \\
\path{tracheal_deviation} & 1,271 & 732 & 310 & 0 \\
\path{tracheostomy_tube} & 1,557 & 4,414 & 5,165 & 26 \\
\path{whole_lung_or_majority_collapse} & 101 & 373 & 171 & 1 \\
\end{longtable}
\endgroup